\documentclass[proceedings, preprint]{rmaa}

\usepackage{paralist}
\usepackage{natbib}

\usepackage{psfrag,color}

\usepackage[latin1]{inputenc}

\def\ms{\mbox{$M_{\rm s}$}}
\def\mh{\mbox{$M_{\rm h}$}}

\def\Mtran{\mbox{$M_{\rm migr}$}}

\def\fs{\mbox{$F_{s}$}}
\def\fb{\mbox{$F_{b,U}$}}
\def\mstar{\mbox{$M^{*}$}}

\def\msun{\mbox{$M_{\odot}$}}
\def\lcdm{\mbox{$\Lambda$CDM}}
\def\gsmf{\mbox{$GSMF$}}
\def\gsmfs{\mbox{$GSMF$s~}}

\def\lesssim{{_ <\atop{^\sim}}}
\def\grtsim{{_ >\atop{^\sim}}}

\SetYear{2011}
\SetConfTitle{XIII IAU LARIM}

\title{The dark and stellar mass assembly of galaxies} 

\author{
    V.~Avila-Reese,\altaffilmark{1} and  C.~Firmani\altaffilmark{1,2}}

\altaffiltext{1}{Instituto de Astronom\'ia,
  Universidad Nacional Aut\'onoma de M\'exico, A.P. 70-264, 04510 M\'exico
  D. F., M\'exico (Email: avila@astro.unam.mx).}
\altaffiltext{2}{INAF-Osservatorio Astronomico di Brera, via E.Bianchi 46, I-23807 Merate, Italy}

\shortauthor{Avila-Reese \& Firmani}
\shorttitle{The stellar and dark halo mass assembly of galaxies}

\listofauthors{V. Avila-Reese, \& C. Firmani}
\indexauthor{Avila-Reese, V.}
\indexauthor{Firmani, C.}

\abstract{The emerging empirical picture of galaxy stellar mass (\ms) assembly shows that
galaxy population buildup proceeds from top to down in \ms.  By connecting
galaxies to \lcdm\ halos and their histories, individual (average) \ms\ growth tracks can be inferred.
These tracks show that massive galaxies assembled their \ms\ the earlier the more massive
the halo, and that less massive galaxies are yet actively growing in \ms, the more active the
less massive is the halo. The predicted star formation rates as a function of mass and the downsizing
of the typical mass that separate active galaxies from the passive ones agree with
direct observational determinations. This implies that the \lcdm\ scenario is consistent with these
observations. The challenge is now to understand the baryonic physics that drives the significant 
and systematical shift of the stellar mass assembly of galaxies from the mass assembly of their 
corresponding halos (from halo upsizing to galaxy downsizing). }

\resumen{}

\addkeyword{Cosmology: theory}
\addkeyword{Galaxies: formation}
\addkeyword{Galaxies: evolution}

\begin{document}

\maketitle

\section{Introduction}
\label{sec:intro}

Galaxies are fascinating astronomical systems. On the one hand, 
the complexity of stellar populations, ISM, AGN,
and dark matter is integrated inside them. On the other hand, 
galaxies are the lost link between the early universe and the observed
world of astronomical objects; as such they trace the way cosmic 
structure emerged in the universe and are the basis of unique methods 
for determining several cosmological parameters.  
One of the current challenges in the study of galaxies is determining
and understanding how did they assemble their stellar and dark halo 
masses. The main observational imprint of the complex galaxy assembling 
processes is probably the stellar-to-dark mass ratio, $\fs\equiv \ms/\mh$, as a function
of mass and redshift $z$.  

The growth of \ms\ can be due (1) to local star formation (SF) fed by available
or infalling gas and/or (2) to capture of other stellar systems.
In the former case, SF is driven by internal physics, mainly by
self-regulation processes in the galaxy disk \citep[e.g.,][]{FT92,FA00}, or is
induced by external interactions and mergers --mainly those with 
high gas content. In the latter 
case, \ms\ can grow by accretion of small satellites and tidal 
debris or by major mergers. Even more, the mentioned processes
imply feedback effects, like SN-driven outflows and AGN-driven intrahalo 
gas reheating/ejection effects that inhibit further \ms\ growth.
{\it Which of these channels and their respective
feedback effects did dominate in the \ms\ growth of galaxies as a function
of their halo masses, types, and environment? How does proceed the stellar and
dark mass assembly of galaxies?}

\section{Different approaches to galaxy formation and evolution}

\subsection{The inductive or 'archaeological' method}
The approaches for studying galaxy formation and evolution have 
changed over time. In the past, the main approach
was the 'archaeological' method: ages and star formation rate
histories of our and other local galaxies are reconstructed by means of stellar population
synthesis and chemical evolution models from the observed spectro-photometric
and chemical properties. The application of this method
to modern data and large galaxy surveys have allowed to 
obtain key constraints to galaxy mass assembly as a function of
mass, galaxy type, and environment
(e.g., Thomas et al. 2005; Panter et al. 2007; Asari et al. 2007). 

\subsection{The deductive or ab initio method}
With the consolidation of the popular $\Lambda$ Cold Dark Matter (\lcdm) cosmology 
in the last twenty years, a powerful theoretical background for galaxy formation and evolution
appeared. According to this framework, cosmic structures emerged
from the gravitational evolution of CDM-dominated primordial perturbations.
The distributions, inner properties, and evolution of the collapsed 
CDM structures (halos) were predicted in detail by cosmological N-body simulations.
Ab initio (deductive) models, as well as full numerical simulations including
hydrodynamics, were developed in order to follow the complexity of baryon physics
during the process of galaxy formation and evolution inside the growing 
CDM halos. Several of the physical ingredients and parameters of the models,
as well a some of the sub-grid parameters of the simulations,
are actually determined/calibrated from mainly local observations. The deductive approach
has enormously contributed to our understanding of galaxy formation
and evolution in the cosmological context 
though many questions remain yet unsolved, 
including the validity of the backbone of this approach itself: the existence of the 
elusive CDM. 

The halo mass growth in the \lcdm\ cosmology is on average hierarchical, from bottom 
to up.  How is the \ms\ assembly of galaxies formed inside the CDM halos and how does
it compare with observations?
\textit{A sharp test for the \lcdm\ scenario may arise from this comparison}. For this, 
however, several theoretical and observational "biases" should be first well understood. 
Neither the galaxies trace directly the assembly of their halos --here is implied all the
complex baryon physics--, nor the observed galaxy populations at different redshifts offer 
a direct way to determine the individual evolution of galaxies --instead they are inferred
and the results are strongly dependent on selection effects and biases.
It should be said that misunderstandings and oversimplifications of the scenario have led to 
some incorrect interpretations. For example, it is common to hear that since in the \lcdm\ hierarchical
scenario dark halos assemble through violent major mergers ("walnut tree"), therefore galaxy
formation is merger-dominated, something that could be in conflict with the observed large
abundance of spiral galaxies or the non-negligible fraction of bulgeless galaxies. In fact, most of the 
mass of galaxy-sized \lcdm\ halos (and most of halos) was (were) not assembled by
major mergers but by smooth accretion and minor mergers ("pine tree"; Maulbetsch et al. 
2007; Genel et al. 2010). Besides, only a fraction of halo-halo mergers end producing
galaxy-galaxy mergers. By analyzing the results of the SAM implemented in the Millenium 
simulation (De Lucia et al. 2005), in De Rossi et al. (2009) it was concluded that only $\approx 12\%$  
of Milky Way-type galaxies experienced a major merger during their lifetimes.

\subsection{The empirical approach}
Thanks to the vertiginous advance of instrumental facilities 
and multi-wavelength large-area survey programs, 
the empirical or "look-back time" approach has come to stay. The main
photometric and spectral properties of whole galaxy populations obtained
at different $z$ allow us to infer \ms, SF rate (SFR), and other physical 
properties of galaxies\footnote{For several  of these inferences, stellar population synthesis
models should be used. These models had a vigorous advance in the last 
years, however, several uncertainties remain yet unsolved (for a recent review 
see Bruzual 2010).}. Given the large numbers and high sensitivities of current 
surveys, relatively complete samples  characterized by the mentioned above 
properties down to nearly dwarf galaxy masses and out to high redshifts become
possible (e.g., $\ms\grtsim 10^8$ at $z\approx0$ or $\ms\grtsim 3\ 10^9$ at 
$z\approx2$).
Thus, progress has been made in determining locally and at higher $z$: 
(1) the total galaxy stellar mass function, \gsmf, and its dissection into 
blue/star-forming and red/passive components,
(2) the SFRs of galaxies as a function of their mass traced by different
indicators, from the UV to the radio,
and (3) the galaxy rate of merging at different masses and epochs, 
although with still very large systematic uncertainties.
In some works, these determinations have been obtained as a function of environment.  

Based on all these observational data, a purely empirical picture of 
\ms\ assembly as a function of mass, type and
environment is emerging now.  A general result is that $\sim 50\%$ of the local
stellar mass density was assembled since $z=1$, 
and $\sim 90\%$ since $z=3.5$ (e.g., P\'erez-Gonz\'alez et al. 2008).
Possibly the main new concept distilled from the empirical
picture is that of \textit{cosmic downsizing}, a term coined by \citet{Cowie96} to describe 
the decline with cosmic time of the maximum rest-frame $K-$band luminosity of 
galaxies undergoing active SF. This term has been more recently 
used to describe a number of trends of the galaxy population as a function of mass
that imply in general galaxy buildup from top to down. However, these
different trends are actually related to different astrophysical phenomena and galaxy 
evolutionary stages \citep{Fontanot09}. 
From the most general point of view, the many downsizing manifestations can be separated
 into those that refer to the evolution of: 
 \begin{itemize}

 \item (A) Massive galaxies, which today are on average red and passive (quenched SF).
 Observations show that the high-mass end of the \gsmf\ was in place since high
 $z$ (e.g., Fontana et al. 2004,2006; Drory et al. 2005; Cimatti et al. 2006; Marchesini 
 et al. 2009; Perez-Gonzalez et al. 2008), and evidence a decrease with cosmic time 
 of the characteristic mass at which the SFR is dramatically quenched or at which the \gsmfs\
 of early- and late-type galaxies cross, i.e. less and less massive migrate with time to the
 red sequence (e.g., Bundy et al. 2006;  Borch et al 2006; 
 Bell et al. 2007; Hopkins et al. 2007; Drory \& Alvarez 2008; Vergani et al. 2008; Pozzetti et al. 2010). 
 This is in line with 'archaeological' inferences for local galaxies that evidence an early and
 coheval mass assembly for massive ellipticals \citep[e.g.,][]{Thomas+2005}. 

\item (B) Less massive galaxies, which are on average blue and star-forming.
Although the observations of less luminous galaxies suffer incompleteness as $z$ increases
due to the flux limits, at least up to $z\sim 1-2$, most of observational studies have 
found that the specific SFR (SSFR = SFR/\ms) of galaxies with $\ms\lesssim 3\times 10^{10}$ \msun, which are mostly 
late-type, star-forming galaxies, is surprisingly high even at $z\sim 0$ (late mass assembly)
and, on average, the lower the mass, the higher the SSFR \citep['downsizing in SSFR'; e.g.,][]{B04,Bauer05,Zheng07,Noeske07,Bell07,Elbaz07,Salim07,Chen09,Damen09b,
Santini09,Oliver10,Kajisawa10,Rodighiero10,Karim10,Gilbank11}. 

\end{itemize}

\begin{figure*}[!t]
\vspace{8.cm}
\includegraphics{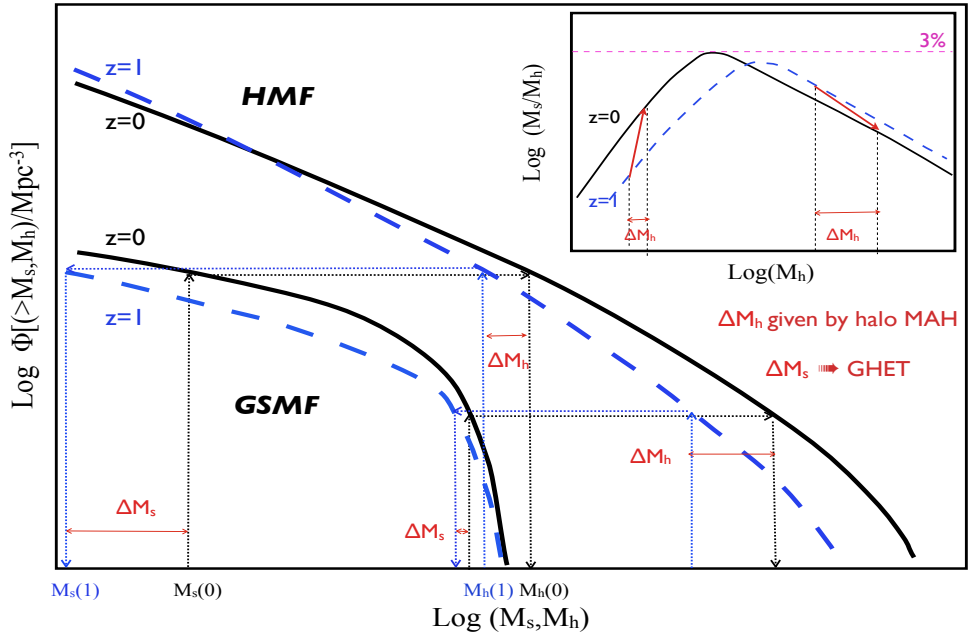}
  \caption{Halo and galaxy stellar mass cumulative functions at $z=0$ (solid black
  curves). Given a \ms, its corresponding \mh\ is found by matching the mass
  functions (the dotted black arrows show this for two masses). A relation between
  \ms\ (or \ms/\mh) and \mh\ is constructed this way (solid black curve in the 
  inset). The same exercise can be repeated at other redshifts (e.g., $z=1$, dashed
  blue curves). Given the halo average MAH, the mass at $z=1$, \mh(1),
  that had a halo of a given mass at $z=0$, \mh(0), can be calculated ($\Delta \mh$).
  Then, by  using again the AMT, the \ms(1) corresponding to \mh(1) is obtained 
  (the dotted blue arrows show this for two masses), and $\Delta \ms =\ms(0)- \ms(1)$ 
  gives the \textit{individual} average \ms\ track (GHET; the red arrows in the inset show
  two GHET-based tracks in the \ms/\mh--\mh\ diagram).   
  }
  \label{scheme}
\end{figure*}

By combining the evolution of the \gsmf\ with the measured SFR--\ms\ relations at
different $z$, the contribution of local SF and galaxy accretion (mergers) to the stellar mass build 
up can be constrained (Bell et al. 2007; Drory \& Alvarez 2008; Pozzetti et al. 2010). 
These and some direct --but yet limited-- observational determinations of the merging
rate (e.g., Lotz et al. 2007; Bundy et al. 2009) show that the former channel completely dominates in low- and 
intermedium-mass galaxies at all epochs, while (dry) mergers may play a moderate role 
for massive ($\ms\grtsim 10^{11}\msun$, mainly red) galaxies at later epochs 
($z\lesssim 1$). 

Although important efforts have been made already in order to constrain the 
dynamical evolution of galaxies (e.g., from observational studies at high redshifts 
of the Tully-Fisher and Faber-Jackson relations, galaxy-galaxy weak lensing, etc.),
direct constraints of the galaxy-halo connection as a function
of mass are yet very limited. This kind of observations together with those regarding
the gas mass of galaxies at high $z$ are necessary to complete the whole
picture of galaxy stellar, baryonic, and dark mass assembly.  In the meantime,
semi-empirical methods were introduced in order to get the
whole evolutionary connection in the context of the \lcdm\ scenario
and constrain in this way the average \textit{individual} trends of the galaxy evolution 
process as a function of mass (e.g., Conroy \& Wechsler 2009; Hopkins et al. 2009; 
Firmani \& Avila-Reese 2010).

\section{Connecting galaxies to dark halos}

In the semi-empirical approach, the information provided by direct observations 
at different epochs is combined in a statistical way with predicted 
properties for the \lcdm\ halos in order: (1) to attain a connection between 
galaxies and halos, and (2) to infer \textit{individual} galaxy 'hybrid' evolutionary
tracks (GHETs).  An schematic idea of the approach 
is shown in Fig. \ref{scheme} and explained below:

{\bf (1) Global galaxy-halo connection.-} 
Given the measured cumulative \gsmf\ and the cumulative halo mass function, $HMF$
(both at the same epoch e.g., $z=0$, black solid curves),
and assuming a one-to-one correspondence between \ms\ and \mh,
both functions are matched, $\Phi_g(>\ms) = \Phi_h(>\mh)$, for finding the \mh\ corresponding
to a given \ms\ (dotted arrows, for two different masses, in Fig. \ref{scheme}), i.e. 
the \ms--\mh\ or $\fs\equiv \ms/\mh-\mh$ relations are inferred (solid bell-shaped curve
in the inset). This method, called the abundance matching technique (AMT;
e.g., Vale \& Ostriker 2004; Kravtsov et al. 2004; Behroozi, Conroy \& Wechsler 2010 -BCW10-, and
for more references see therein), makes a minimum of assumptions and has proven 
to be effective and practical.  The halo masses of galaxies can be actually determined
by direct methods as galaxy-galaxy weak lensing and satellite kinematics (e.g., 
Mandelbaum et al. 2006; More et al. 2010). Nevertheless, in current studies, due to
the low signal-to-noise ratios, results can be obtained only by stacking a large
number of galaxies; this introduces biases and significant statistical uncertainties
in the inferences, and limits these inferences to relatively small mass ranges.
However, in the mass ranges where comparisons are feasible, these direct
methods, the model-dependent indirect methods (e.g., the Halo Occupation Model and the Conditional
Luminosity Function formalism), and the AMT, give local \fs--\mh\ relations compatible
among them within a factor of $\sim 2$ (see BCW10; Moster et al. 2010; Guo et al. 2010;
More et al. 2010; Rodr\'iguez-Puebla et al. 2011).  

\begin{figure*}[!t]
\vspace{8.5cm}
\includegraphics{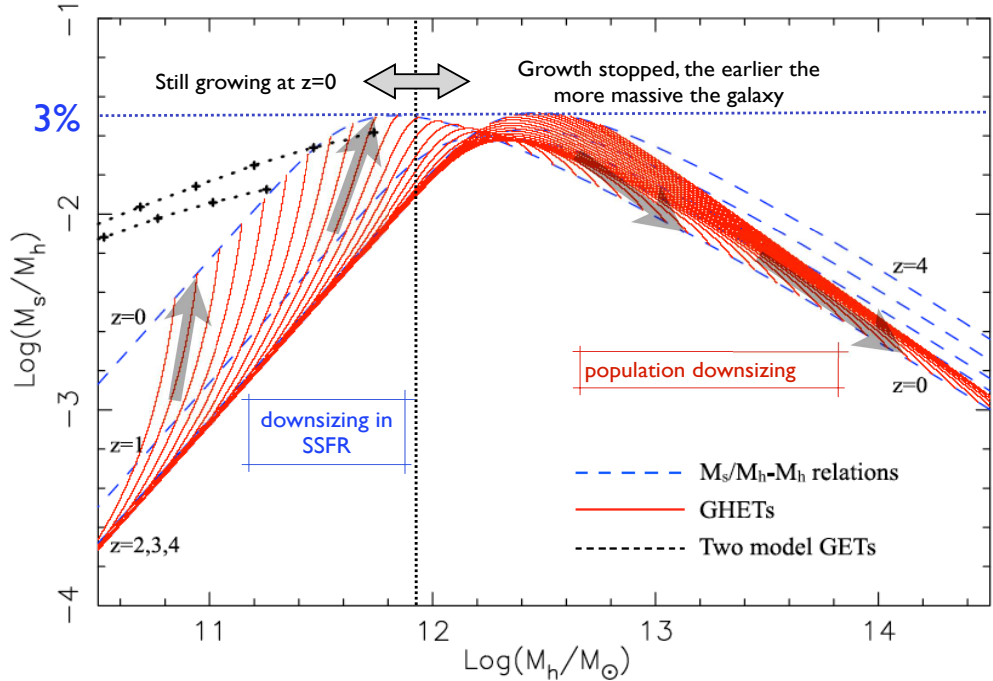}
  \caption{The \ms/\mh--\mh\ functions inferred by the AMT at four $z$ (dashed
  blue curves; BCW10, see FA10 for modifications).  Solid red curves are
  the average \ms\ growth tracks inferred in FA10: in small halos, \ms\ assembles
faster at late epochs as smaller is \mh, while in large halos, \ms\ stopped
its growth, the earlier the more massive the halo. The two dotted curves
are tracks corresponding to two disk galaxy evolutionary models.
 }
  \label{fs-Mh}
\end{figure*}

The obtained stellar mass fractions, \fs, are very low with respect to the universal baryon
fraction ($\fb\equiv \Omega_b/\Omega_M\approx 0.16$), i.e. galaxy SF inside \lcdm\ halos
seems to be a very inefficient process. Besides, the efficiency is strongly dependent on mass:
it peaks around $\mh=8\ 10^{11}$ \msun, decreasing significantly towards lower ($\fs\propto \mh^{a},
a\approx 1.25$ ) and higher ($\fs\propto \mh^{b}, b\approx -0.6$) masses.
The \gsmf\ determinations at higher redshifts allow now to use methods like the AMT to infer the 
\ms--\mh\ relation at different epochs \citep[][BCW10]{Moster10,Wang10}. The latter authors 
extended the AMT out to $z\approx 4$ and found that (a) the \mh\ at wich \fs\ peaks shifts 
little to higher masses (by $\approx 0.6$ dex out to $z\approx 4$), and (b) the peak value remains 
roughly constant. For masses below (above) the peak at a given epoch, \fs\ is smaller (slightly larger) 
than at earlier epochs for a given \mh\ (see Fig. \ref{fs-Mh}, dashed blue curves). This 
implies that galaxy SF in massive halos should have been slightly more efficient in the past, while
in low-mass halos it should have been even less efficient. {\it This is a manifestation of the cosmic
downsizing discussed above. }

{\bf (2) Individual \mh\ and \ms\ growth tracks.-}  In the \lcdm\ scenario,
the individual mass aggregation histories (MAHs) of halos are known. By using them in
combination with the semi-empirical \ms--\mh\ relations at different $z$, the corresponding \ms\ evolutive
tracks (GHETs) can be inferred \citep[red arrows in the inset of Fig. \ref{scheme};][hereafter FA10]{CW09,FA10}. 
In FA10, average halo MAHs and a parametrization of the BCW10 \ms--\mh\ relations
up to $z=4$ were used to infer individual \textit{average} GHETs.  The
results are encouraging and show that at each epoch there is a characteristic 
mass that separates galaxies into two populations (Fig. \ref{fs-Mh}): (a) galaxies more 
massive than \ms($z=0$)$\approx 3\times 10^{10}$ \msun\ are on average quiescent/passive 
(their \ms\ growth slowed down or stopped completely), besides the more massive
 is the galaxy, the earlier it has transited from the active (blue, star-forming) to the passive
 (red, quenched) population ('population downsizing'); (b) galaxies less massive than 
 \ms($z=0$)$\approx 3\times 10^{10}$ \msun\ are on average active (blue), and the 
 less massive the galaxy, the faster its late \ms\ growth, driven likely by local SF
 ('downsizing in SSFR').  
 
\begin{figure}[t!]
\vspace{6.65cm}
\includegraphics{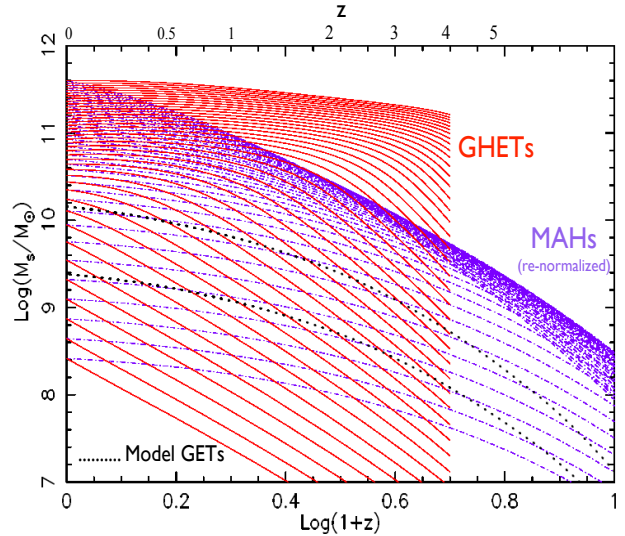}
  \caption{Average \ms\ growth tracks (GHETs) as a function of $z$
  (solid red curves) and their corresponding halo MAHs (dot-dashed 
  purpure curves) shifted vertically
  in order \ms($z=0$)=\mh($z=0$). Galaxies and halos assemble
  their masses in a very different way (from FA10).}
  \label{GHETs}
\end{figure}

 The \ms($z$) average tracks (GHETs) inferred in FA10 are shown in Fig. \ref{GHETs} (solid red curves).
Their corresponding halo MAHs are also plotted (dot-dashed blue curves) but for comparative 
reasons, they were shifted vertically in such a way that each MAH coincides with its related
GHET at $z=0$. The shapes of the average stellar and halo mass assembling histories are 
quite different. 
For galaxies with $\ms(z)< 3\ 10^{10}$ \msun, their halo MAHs at later epochs 
grow slightly slower as smaller is the mass, while their stellar GHETs grow much faster. 
For \ms($z$=0)$ > 3\ 10^{10}$ \msun, as the system is more massive, the stellar assembly 
of the galaxy occurs earlier in time with respect to the corresponding halo. Only for systems
that attained at a given epoch masses $\ms\approx 3-5\ 10^{10}$ \msun\ 
($\mh\approx 1-2\ 10^{12}$ \msun), both the \ms\ and \mh\ assembly history shapes are similar; 
these systems are namely those in the peak efficiency (maximum \ms/\mh\ ratio, see Fig. \ref{fs-Mh}). 

 \subsection{Predictions}
 
\begin{figure}[!t]
\vspace{6.65cm}
\includegraphics{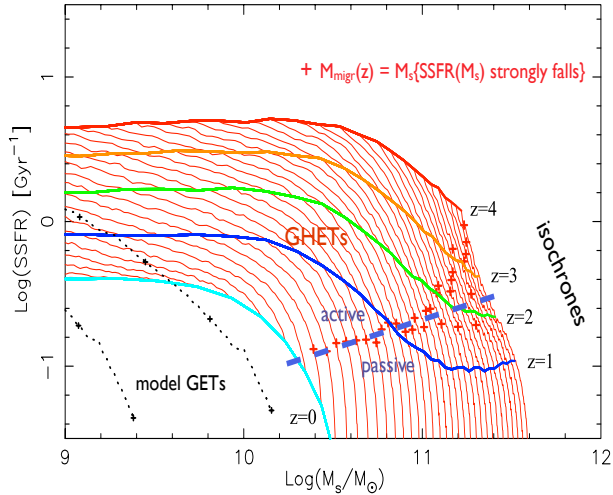}
  \caption{Predicted average SSFR--\ms\ tracks (assuming negligible
  \ms\ growth by mergers; solid thin red curves, from FA10). The thick solid
  curves connect the individual tracks at a given $z$ (isochrones).
  The crosses show when a given track strongly falls --the galaxy
  becomes passive. The mass \ms\ at this epoch is called \Mtran.
  }
  \label{ssfr}
\end{figure}

 {\it SSFR histories.-} 
By using the GHETs, the specific \ms\ growth rate histories, $\dot{\ms}(z)/\ms(z)$, 
 can be calculated. If as a working hypothesis one assumes that the \ms\ growth of a given
 galaxy is only due to local SF, then its SSFR history is $\dot{\ms}(z)/\ms(z)$ 
 divided by $(1-R)$, where
 $R\approx 0.4$ is the gas recycling factor due to stellar mass loss. In Fig. \ref{ssfr} we 
 reproduce the individual average SSFR tracks in the SSFR vs \ms\ diagram as obtained in 
 FA10 (thin red solid lines): the SSFR of low-mass galaxies decreases with time
 slowly, their late \ms\ growth being very fast, while for large galaxies, the SSFR
 decreases very fast and their \ms\ growth practically stops at late epochs. 
 
 The thick solid lines correspond to isochrones at $z=0, 1, 2, 3$ and 4, from bottom 
 to top, respectively. These isochrones are the SSFR--\ms\ relations at the given $z$
 and can be compared with the direct determinations mentioned in \S\S 2.3. Within the current
 large uncertainties and sample selection effects, the GHET-based and directly determined
 average SSFR--\ms\ relations out to $z\sim 2$ are in reasonable agreement. If any, the former
 are shallower than the latter at the low-mass side; at low $z$, the predicted SSFRs for intermediate
 masses ($\ms\sim 0.7-2\ 10^{10}$ \msun) are slightly larger than the averages of those directly 
 measured. It is important to note that the fast decreasing in the SSFR--\ms\ relations (isochrones)
 seen at  the high-mass side is associated to passive (red) galaxies. Most of the SSFR--\ms\ relations 
 presented in the literature refer to only star-forming (blue) galaxies; for masses lower than
 $\ms\approx 2-3\ 10^{10} \msun$, blue galaxies by far dominate in number, but at high masses
 passive (red) galaxies become dominant (e.g., Bell et al. 2003; Salim et al. 2007; Pozzetti et al. 2010). 
 In the works where the samples were partitioned by color, the SSFR--\ms\ relation of the red galaxies
 is inferred indeed very steep \citep[e.g.,][]{Bell07,Karim10}.

\begin{figure}[!t]
\vspace{6.2cm} 
\includegraphics{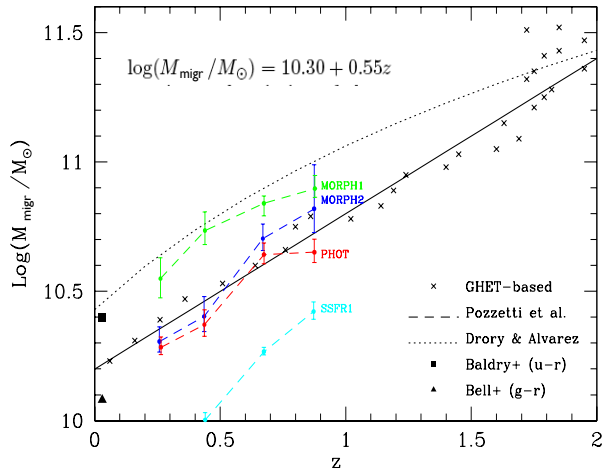}
  \caption{Typical mass at which galaxies migrate on average to the passive
  sequence at a given $z$ (crosses, see Fig. \ref{ssfr}). The solid
  line is a eye-fit to the crosses. The dashed lines correspond to
  the mass $M_{\rm cross}$ at which the late- and early-type
  \gsmfs\ crosses, where different morphological, photometrical
  and SSFR criteria were used to separate the sample into
  late- and early-type galaxies (Pozzetti et al. 2010). See text
  for more details. 
  }
  \label{Mtrans}
\end{figure}

 {\it  Downsizing of the migration (quenching) mass.-} 
 The GHET-based SSFRs can be used to calculate the $z$ at which the SSFR as a function
 of \ms\ for a given track decreases dramatically (the galaxy migrates to the
 passive sequence). The crosses in Fig. \ref{ssfr} show, for each track, the \ms\ 
 attained by the galaxy at this epoch, i.e. at each epoch there is a mass \Mtran\
 at which galaxies on average came to be passive (Fig. \ref{Mtrans}).
 The solid line in Fig. \ref{Mtrans} is a by-eye linear fit to the data: 
log$(\Mtran/\msun) = 10.30 + 0.55z$, i.e. the population migration progresses
from massive systems at high $z$ to less massive systems at lower $z$ ('population
downsizing').  Is there evidence from direct observations
of a migration mass to the passive population as a function of $z$? 
As mentioned in \S\S 2.3, observations allow now to determine the
\gsmf\ at different epochs decomposed into early-- (red) and late--type (blue) galaxies.
For example, based on the $z$COSMOS survey, Pozzetti et al. (2010) determined 
the mass at which the early-- and late--type \gsmfs\ cross (different estimators 
for these two populations are used) from $ z\approx 1$ to $z\approx 0.2$. 
Such a crossing mass, $M_{\rm cross}$, is interpreted namely as the typical mass of late-type 
galaxies migrating to early-type ones (see also Bell et al. 2007). Local estimates of 
$M_{\rm cross}$ by Bell et al. (2003; filled triangle) and Baldry et al. (2004; filled square) 
are also plotted, as well as the law inferred by Drory \& Alvarez (2008) for the mass above 
which the SFR as a function of \ms\ begins to drop exponentially, $M_{\rm quench}/\msun$ = 
$10^{10.43}$($1+z$)$^{2.1}$. It is remarkable the agreement between
direct estimates of different definitions of the characteristic mass above which the 
passive (red) population of galaxies dominates in number density (see also Bundy 
et al. 2006; Hopkins et al. 2007; Vergani et al. 2008) and the semi-empirical 
inferences of FA10.

{\it Migration rates and the quasar connection.-} 
Another prediction that can be made with the semi-empirical approach is the
rate in comoving number density of galaxies migrating from active to passive ones 
as a function of $z$.  In FA10 it was found that this rate, $\dot{\phi}_g(\Mtran)$, 
up to $z=1$ scatters around $(1.0 \div 5.5)\times 10^{-4}$ gal/Gyr/Mpc$^3$ without 
any clear trend with $z$. These rates are in agreement with the estimated passive (red)
population growth rate per unit of comoving volume determined by Pozzetti 
et al. (2010) for a redshift bin centered at  $z=0.34$. 

\begin{figure}[!t]
\vspace{6.1cm}
\includegraphics{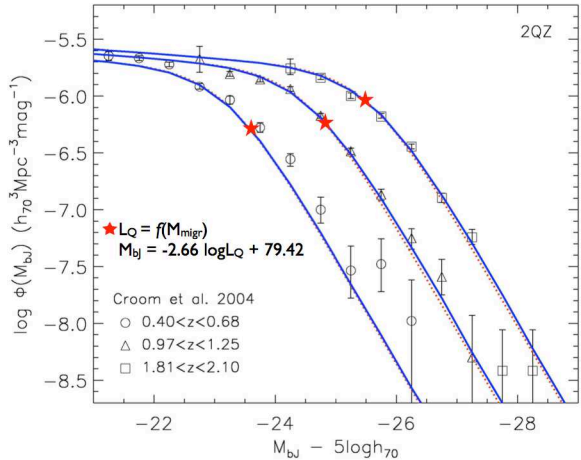}
  \caption{The \Mtran\ mass and its abundance translated to a quasar
  luminosity, $L_Q$, and abundance at three redshifts ($z=0.5, 1,$ 
  and 2, filled stars).  The obtained results are showed over the luminosity
  function plots of quasars at the three $z$ bins indicated in the panel
  (Croom et al. 2004; Croton 2009; see text for details). The masses and 
  abundances of galaxies in the migration process roughly coincide with the characteristic
  $L_Q$ of the quasar luminosity functions at the same redshifts.}
  \label{Lq}
\end{figure}

If the typical time of galaxy transition (or quenching) is $t_Q$, then the abundance
of galaxies per unit of comoving Mpc in the process of migration, i.e. those of
mass $\Mtran$, is $\phi_{\rm g,migr}\approx \dot{\phi}_g(\Mtran)\times t_Q$. Let assume 
$t_Q=100$ Myr and let estimate the typical luminosity of a bright quasar, $L_Q(z)$ 
associated to the stellar mass $\Mtran(z)$ or its corresponding halo mass \mh($z$) 
(see e.g., Croton 2009) by making some assumptions ($L_Q=\eta L_{\rm Edd}$, 
 $L_{\rm Edd} = C\times M_{\rm BH}$, $M_{\rm BH}$ is the supermassive black hole mass) 
 and by using the observable correlations of 
 $M_{\rm BH}$ with spheroid velocity dispersion or stellar mass. The result is very 
encouraging (Fig. \ref{Lq}): at each $z$ (up to $z\sim 3$), both
$L_Q(\Mtran)$ and its abundance, $\phi_{\rm g,migr}$, match the characteristic luminosity 
of the two-power law QSO luminosity function and its abundance as determined in Croon et al. (2004) 
and Roberts et al. (2006). This suggests that the transition from active to passive regime 
(quenching) in massive galaxies is associated to the QSO active phase.

\subsection{Implications}
 
 In the semi-empirical approach described above, the observational input is the total
 \gsmf\ measured at different $z$, while  the information provided by the \lcdm\ theory
 allowed to assign halos to galaxies and "individualize" average galaxy \ms\ tracks 
 as a function of mass. By using the 
 inferred \ms\ tracks (GHETs), the SSFRs as a function of \ms\ and $z$, the 
 evolution of the migration or quenching mass \Mtran, and the rate per unit of 
 comoving volume of galaxies transiting from active to passive ones were predicted. 
 The fact that these predictions are in reasonable agreement with direct observational
 measures at different $z$ implies that {\it both the connection between
 galaxies and \lcdm\ halos, and the predicted halo MAHs are consistent with
 observations.} 
  
The semi-empirical results imply in a natural way the bimodal distribution of galaxies 
into active (blue) and passive (red) ones, with a bimodality mass scale that shifts downward
as time progress (more galaxies become on average passive), this mass scale being at $z=0$ 
$\ms= 2-3\times 10^{10}$ \msun, in agreement with direct observations (e.g., Bell et al. 2003;
Kauffmann et al. 2003; Blanton \& Moustakas 2009). 
Therefore, {\it mass seems to be the main driver of color, SFH, and SF quenching,} 
though environment is expected to play also a role. Observational studies
show indeed that color and SFR 
are moderately modulated by environment for intermediate-mass galaxies,
which assembled more recently, in epochs when larger-scale
structures collapse, affecting this those galaxies living in such overdense
regions (Cucciati et al. 2010; Peng et al. 2010). The downsizing of the quenching
mass has been found also to be slightly accelerated in overdense regions (Bundy et al. 2006).

The inference of SSFR (as well as \Mtran) as a function of $z$ was under the hypothesis that the
\ms\ growth is driven only by local SF, i.e. the growth by accretion of stellar systems (dry mergers) was
excluded. The rough agreement between predictions and direct 
measurements implies that this hypothesis is reasonable or, said in other words,
that the former growth channel dominates. 

\section{Challenges and concluding remarks}

The semi-empirical results imply that {\it galaxy formation
inside the growing \lcdm\ halos is consistent in general with observations}. 
However, these results show that the average stellar mass assembly histories 
of galaxies shift from those of their halos in a peculiar way and depending on 
mass (Fig. \ref{GHETs}). The complex astrophysical processes and environmental 
effects involved in galaxy evolution should explain this. Among the key
challenges, we highlight and discuss the following ones:

\subsection{Why the \ms\ assembly of low-mass galaxies is systematically delayed with
respect to the assembly of their halos?}

The SAMs introduce (extremely efficient) SN-driven galaxy outflows for reproducing 
the shallow faint-end of the luminosity function or, equivalently, the high inclination of 
\ms/\mh\ at lower masses (Fig. \ref{fs-Mh}).  Can galaxy outflows explain also the
inefficient galaxy SF in the past as smaller are the halos? By means of
self-consistent disk-galaxy evolutionary models, Firmani, Avila-Reese
\& Rodr\'iguez (2010) have shown that SN-driven outflows, as well as the
local SF and ISM feedback processes, may deviate the SFR history of disk galaxies
from the associated halo mass aggregation rate history but not enough
as to reproduce the semi-empirical and empirical results. In Figs. \ref{fs-Mh}, \ref{GHETs}, 
and \ref{ssfr} two evolutionary models of low-mass disk galaxies are plotted 
(dotted-line curves). The model tracks differ significantly from the GHETs. As seen
in Fig. \ref{GHETs}, the model \ms\ tracks follow closely those of their
halos at late epochs, in disagreement with the systematical 
shift as the mass is smaller observed for the corresponding GHETs. Although the low \ms/\mh\
ratios were obtained in the models at $z=0$ (by assuming very high SN kinetic-energy 
injection to the outflow, see also e.g., Dutton \& van den Bosch 2009), at higher $z$, 
model prediction are far from the semi-empirical inferences (Fig. \ref{fs-Mh}).
The disagreement between models and observations is also evident by comparing the 
SSFRs:  the former predict SSFRs for small galaxies much lower than the
observed ones, specially toward lower $z$.  

Firmani et al. (2010) explored the possibility of later re-accretion of the 
ejected gas (e.g., Oppenheimer et al. 2010). For reasonable schemes of gas 
re-accretion as a function of \mh, they found that the SSFR of galaxies 
increases but in the opposite direction of the downsizing trend:  the 
increase is large for massive galaxies and small for the less massive ones.
From the side of SAMs, it was found that the population of small galaxies (both central
and satellites) is too old, red, and passive as compared with observations 
(e.g., Somerville et al. 2008; Fontanot et al. 2009; Santini et al. 2009; Liu et al. 2010),
issues that certainly are related to the mentioned-above problem of the \ms\ buildup 
of \lcdm-based sub-\mstar\ model galaxies.  Current N-body + hydrodynamics cosmological
simulations (where feedback-driven outflows are allowed) of central low-mass galaxies ($0.2\lesssim \ms/10^{9} 
\msun \lesssim 30$ at $z=0$) show similar problems (Col\'in et al. 2010; Avila-Reese et al. 2011): 
(1) the SSFRs are 5-10 times lower than the average of observational determinations at $z\sim 0$,
an inconsistency that apparently remains even at $z\sim 1-1.5$, though less drastic;
(2) the \ms/\mh\ ratios are $\sim 5$-10 times larger than observational inferences at $z\approx0$ and 
this difference increases at higher $z$.  

Unless the current observational inferences of \ms, SFR, and their distributions at different $z$ 
are dominated by strong systematic effects and selection biases (see for discussions BCW10; 
Firmani et al. 2010; Avila-Reese et al. 2011), {\it the problems highlighted above pose 
a sharp challenge to current \lcdm-based models and simulations of low-mass galaxies}. Its solution 
requires that galaxies smaller than \ms($z=0$)$\sim 2\times 10^{10}$ \msun\ should have
significantly delayed their \ms\ assembly with respect to the assembly of their corresponding \lcdm\ halos,
besides the smaller the galaxy, the longer should be such a delay.
This is in line with the staged galaxy formation picture proposed in \citet{Noeske07}.

\subsection{Why the more massive the galaxy, the earlier  assembled most of its \ms?} 
Though at first glance contradictory to the \lcdm\ picture, this manifestation of downsizing
has, at least partially, its natural roots namely in the hierarchical clustering assembly of 
halos and their progenitors distributions \citep[][see also Guo \& White 2008; Li, Mo 
\& Gao 2008; Kere{\v s} et al. 2009]{Neistein06}.  Besides, red massive galaxies are 
expected to have been formed in early collapsed massive halos--associated to high peaks,
therefore highly clustered-- that afterwards become part of groups 
and clusters of galaxies, leaving truncated therefore the early efficient mass growth of the 
galaxies associated to these halos. The measured correlation function of luminous red galaxies
is indeed very high (e.g., Li et al. 2006).  On the other hand, massive galaxies typically 
hosted in the past AGNs. The strong feedback of the AGN may 
help to stop gas accretion, truncating further the galaxy stellar growth and giving rise to 
shorter formation time-scales for the massive galaxies \citep{Bower06,Croton06,DeLucia06}.
Although all these effects may explain the archeological downsizing, several questions
remain yet open and subject  to observational testing, among them, whether 
AGN-feedback is as efficient as required for quenching galaxies. 

\subsection{Why galaxies transit from active to passive \ms\ growth regimes, the more massive
earlier than the less massive ones? What does quench the SFR of galaxies?}
These questions are related to the previous one but make emphasis on the 
continuous and monotonic decreasing of \Mtran\ with time.  An important question
to take into account is the possible link between SFR quenching and morphological
transformation; after all, active (blue) and passive (red) galaxies use to be identified
with disk- and spheroid-dominated systems, respectively.
As numerical simulations show, the main driver of morphological evolution seem
to be mergers  (e.g., Scannapieco \& Tissera 2003; Cox et al. 2006 and more references therein). 
Mergers can also quench SF through the associated explosive quasar or starburst phase that heats or 
drive out cold gas in the spheroidal merger remnant. Thus, the scenario where
\lcdm-based mergers drive the formation and quenching of red, early-type galaxies passing 
by a quasar and/or starburst phase looks promising (Kauffmann \& Haehnelt 2000; 
Hopkins et al. 2008, and more references therein).  The connection showed in \S\S 3.1 
between \Mtran($z$) and the characteristic luminosity of the quasar luminosity function at each
$z$ supports this scenario, at least for the formation of red galaxies more massive than
\Mtran\ ($\Mtran\approx 2\ 10^{10}$ \msun\ at $z=0$).  

However, recent observations suggest that the situation is more complicated,
specially for the origin of passive (red) galaxies of smaller masses: nearly 50\% of the
red-sequence COSMOS galaxies ($z\lesssim 1$) have disk-like morphologies,
being bulge-dominated (Bundy et al. 2010), though above $\sim 10^{11}$ \msun, this fraction
is smaller and decreases with time. The authors suggest that passive disks may 
be a common phase of galaxies migrating to the red sequence, and once
formed, their transformation into spheroidals is moderately fast (1--3 Gyr), driven
likely by minor mergers. Passive disks could be the result of disk regrowth
after a gaseous (wet) major merger (Springel \& Hernquist 2005; Governato
et al. 2009), where the formed bulge may host an AGN able to quench further
SF or the bulge can act as a suppressor of disk instabilities and SF (Martig et al. 2009).
In Bundy et al. (2010) are explored several alternatives in the light of their
observational results; they conclude that likely some combination of processes
and environmental effects may be operating  simultaneously in the formation of
the red-sequence galaxies.

\subsection{What are we lacking?}

In case the current observational picture is confirmed with more and deeper observational 
studies, the above-mentioned issues are likely related to our lack of understanding of 
several complex astrophysical and environmental processes of galaxy 
evolution in interaction with the dark mould.  A key ingredient not yet well studied is the
physics of the intergalactic medium (IGM), taking into account the generation and
dissipation of turbulence. The trapping and infall to the center of halos of the IGM, as 
well as the feedback from galaxies (mainly due SNe and AGN) with this medium, 
are relevant phases of galaxy assembly that crucially depend on the physical state of 
the IGM, especially at early epochs. The SF process is another poorly understood ingredient, 
especially in the regime of mergers and strong gas shocks or in subcritical conditions (e.g.,
low gas surface densities). If the above-mentioned issues persist after the 
"bright" side of galaxy formation is well understood in the light of observations, 
then we should have to think about possible modifications to the underlying cosmological 
background, the hierarchical clustering \lcdm\ scenario. 

\vspace{0.5cm}
V.A-R. thanks the organizers for the invitation. We thank PAPIIT-UNAM grant IN114509
for partial funding.

\end{document}